\documentclass[onecollarge,natbib]{svjour2}
\bibpunct{[}{]}{;}{n}{}{,} % to get "[numbered]" references from natbib
\smartqed  % flush right qed marks, e.g. at end of proof
\usepackage{graphicx}
\journalname{Few-Body Systems}

\newcommand{\beq}{\begin{eqnarray}}
\newcommand{\eeq}{\end{eqnarray}}
\newcommand{\nneeq}{\nonumber \end{eqnarray}}

\newcommand{\es}{& = &}

\begin{document}

\title{ Proton structure in high-energy high-multiplicity p-p collisions }
%\thanks{Grants or other notes
%about the article that should go on the front page should be
%placed here. General acknowledgments should be placed at the end of the article.}
% \subtitle{Do you have a subtitle?\\ If so, write it here}
%\titlerunning{Short form of title}        
% if too long for running head

\author{Stanis{\l}aw D. G{\l}azek \and  Patryk Kubiczek }
%\authorrunning{Short form of author list} % if too long for running head
\institute{S. D. G{\l}azek \at
              Institute of Theoretical Physics, Faculty of Physics,
              University of Warsaw, Pasteura 5, 02-093 Warszawa, Poland \\
%              Tel.: +48 22 553 2824\\
%              Fax:  +48 22 553 2995\\
              \email{stglazek@fuw.edu.pl}
           \and
              P. Kubiczek \at
              Faculty of Physics, Astronomy and Applied Computer Science,
              Jagiellonian University, {\L}ojasiewicza 11, 30-348 Krak{\'o}w, 
              Poland \\
              \email{patryk.kubiczek@gmail.com} }

\date{Received: 25 January 2016 / Accepted: 15 March 2016}
% The correct dates will be entered by the editor

\maketitle

\begin{abstract}

A few-body proton image, expected to be derivable 
from QCD in the renormalization group procedure 
for effective particles, is used within the Monte 
Carlo Glauber model to calculate the anisotropy 
coefficients in the initial collision-state of 
matter in high-energy high-multiplicity proton-proton 
interaction events. We estimate the ridge-like 
correlations in the final hadronic state by assuming 
their proportionality to the initial collision-state 
anisotropy. In our estimates, some distinct few-body 
proton structures appear capable of accounting for 
the magnitude of p-p ridge effect, with potentially 
discernible differences in dependence on multiplicity. 

\end{abstract}

%%%%%%%%%%%%%%%%%%%%%%
\section{Introduction}
\label{intro}
%%%%%%%%%%%%%%%%%%%%%%

Protons resist precise theoretical description of 
their internal dynamics in the Minkowski space-time 
for a long time by now. The simplest such picture,
which is provided by the constituent quark model 
used to classify hadrons, is not precisely derived 
from QCD. The theory itself uses the Euclidean-space 
techniques that do not easily yield any real space-time 
image. In these circumstances, it is of interest to 
note that high-energy high-multiplicity proton-proton 
({\it pp}) collisions may shed new light on the issue 
of proton structure. Namely, the numerous products in 
such collisions exhibit collective behavior that appears 
dependent on the initial state of colliding proton 
matter and the latter depends on the proton structure. 
Thus, the correlations among products in high-energy 
high-multiplicity {\it pp} collisions may report 
on the proton structure.

In particular, the CMS~\cite{CMS} and ATLAS~\cite{ATLAS} 
collaborations reported the collective flow in $pp$ 
collisions that resembles the one observed in heavy-ion 
collisions~\cite{bialkowska, li,bozek}. Several authors 
discussed such flows in $pp$ collisions~\cite{dent,casal,
prasad,bozek2,ortona,pierog,bautista,avsar,deng,
v2,v3,bbg,Bjorken} and we follow the insights of Ref.~\cite{dent} 
in order to estimate the extent to which the high-energy 
high-multiplicity $pp$ events are sensitive to the model 
of proton structure. Following the approach of Ref.~\cite{dent} 
and the parameter choice such as in Ref.~\cite{v2}, used here,
means making a strong assumption that the parton medium 
produced in the overlap region of $pp$ collision at the LHC 
has similar hydrodynamical properties as that in heavy 
ion collisions at RHIC. The individual proton structures we 
consider are motivated by the general features of the 
renormalization group procedure for effective particles 
(RGPEP) in quantum field theory~\cite{glazek,gomezglazek}. 

We find that, the effective picture of a quark and diquark 
with a gluon flux between them produces a different dependence 
of eccentricity and triangularity on multiplicity than the 
three-quark picture with a star-like junction made of gluons 
does. According to this finding, the recent data for high-energy 
high-multiplicity events suggest a significant star-like 
gluon junction component in the proton structure. Our analysis 
also indicates a need for assessing the adequacy of the linear 
relationship used by us between asymmetries in the initial 
collision state, such as the eccentricity or triangularity, 
and the final state correlations in high-multiplicity events, 
such as the elliptic flow. 

%%%%%%%%%%%%%%%%%%%%%%%%%%%%%%%%%%%%%%%%%%%%%%%
\section{ Proton structure in $pp$ collisions }
\label{ps}
%%%%%%%%%%%%%%%%%%%%%%%%%%%%%%%%%%%%%%%%%%%%%%%

As mentioned in Sec.~\ref{intro}, the ridge-effect in 
$pp$ scattering can be described using the hydrodynamic 
evolution of the asymmetric state of matter that results 
from one proton's quark and gluon distribution suddenly
colliding with another's. The asymmetric state is meant 
to evolve according to laws of hydrodynamics until it 
eventually turns into the detected particles that emerge 
through hadronization in the final state, in which they 
exhibit the angular correlation over a long-range in 
rapidity, called the ridge. The final state ridge-like 
correlations, such as the elliptic flow, are thus related 
to the initial stage of $pp$ collision whose nature 
depends on the proton structure. One of the key issues 
is thus how to describe the proton structure using QCD
in the Minkowski space-time. 
\begin{figure*}[hb]
\centering
\includegraphics[width=0.8\textwidth]{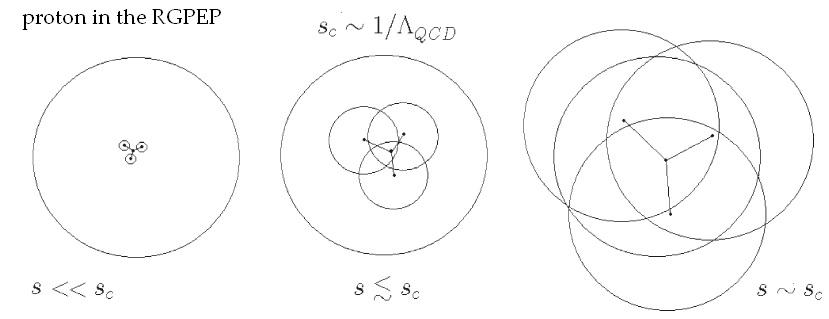}
\caption{Proton structure described using effective
quarks of size $s$ that is very much smaller than, smaller 
and comparable with the constituent quark size $s_c$,
with the single large circle indicating the volume 
available for effective gluons irrespective of their 
corresponding size, see Ref.~\cite{glazek}.}
\label{fig:1}   
\end{figure*}

\begin{figure*}[ht]
 \centering
 \includegraphics[width=0.6\textwidth]{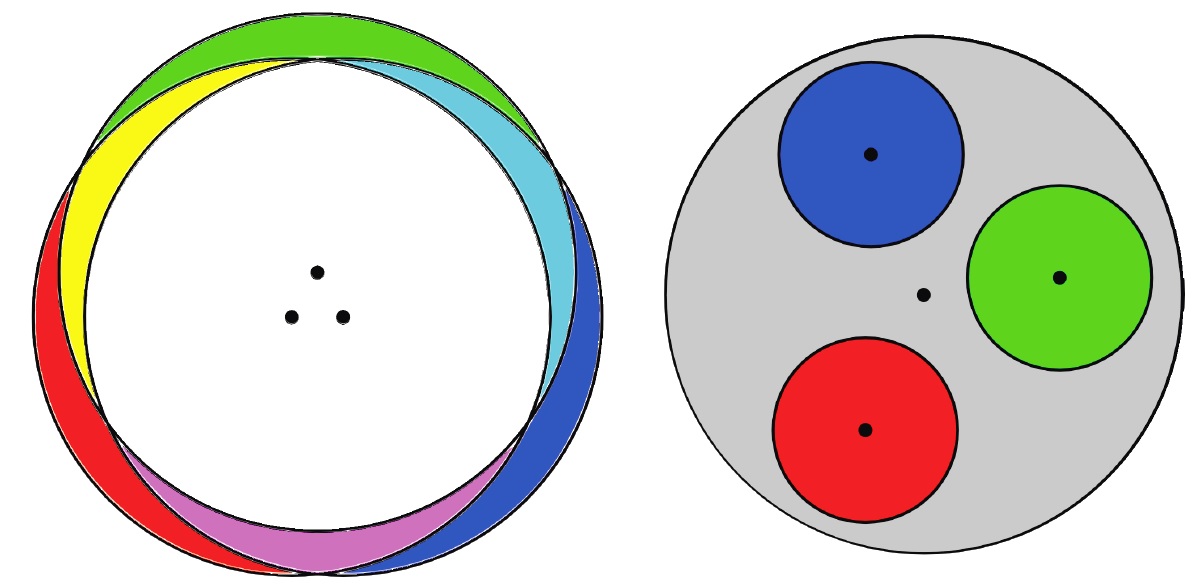}
 \caption{Color structure of effective quarks for two 
 values of the RGPEP scale parameter $s \sim s_c$ and
 $s < s_c$ in Fig.~\ref{fig:1}. Pions are meant to 
 couple to nucleons in the constituent 
 quark picture only at the nucleon boundaries, where color 
 is not neutralized, and for smaller values of $s$ the quarks 
 form more localized objects with the gray area indicating
 the volume available for effective gluons of a similar size
 to the quarks (drawing from Ref.~\cite{Kubiczek}).}
 \label{fig:2}
 \end{figure*}

Conceptually, we approach this issue using the 
RGPEP~\cite{glazek,gomezglazek}, which is a 
candidate for providing the mathematical tools 
for describing protons as bound states of effective 
quarks and gluons of specific size $s$. The size 
parameter $s$ plays the role of an arbitrary 
renormalization-group scale that can be adjusted 
to the physical process one wants to accurately 
describe in simplest possible terms. This condition 
means choosing the right variables for grasping the 
essence of physics most economically from the 
computational point of view. The scale dependence 
of the proton structure expected in the RGPEP is 
illustrated in Fig.~\ref{fig:1} and the corresponding
examples of the color structure are shown in 
Fig.~\ref{fig:2}. 
The expectation is based on
the scale-dependent features of effective Hamiltonians, 
which imply the possibility that a relativistic 
bound-state eigenvalue problem can be equivalently 
written in terms of a few-body problem for sizable 
effective quarks and gluons instead of an infinite 
combination of bare point-like quarks and gluons 
of canonical QCD. Hence, the RGPEP provides the 
scheme in which the distribution of matter in proton 
can be imagined in terms of wave functions, or 
probability distributions for the effective 
quarks and gluons of size $s$.

\begin{figure}[hb]
 \centering
 \includegraphics[width=0.4\textwidth]{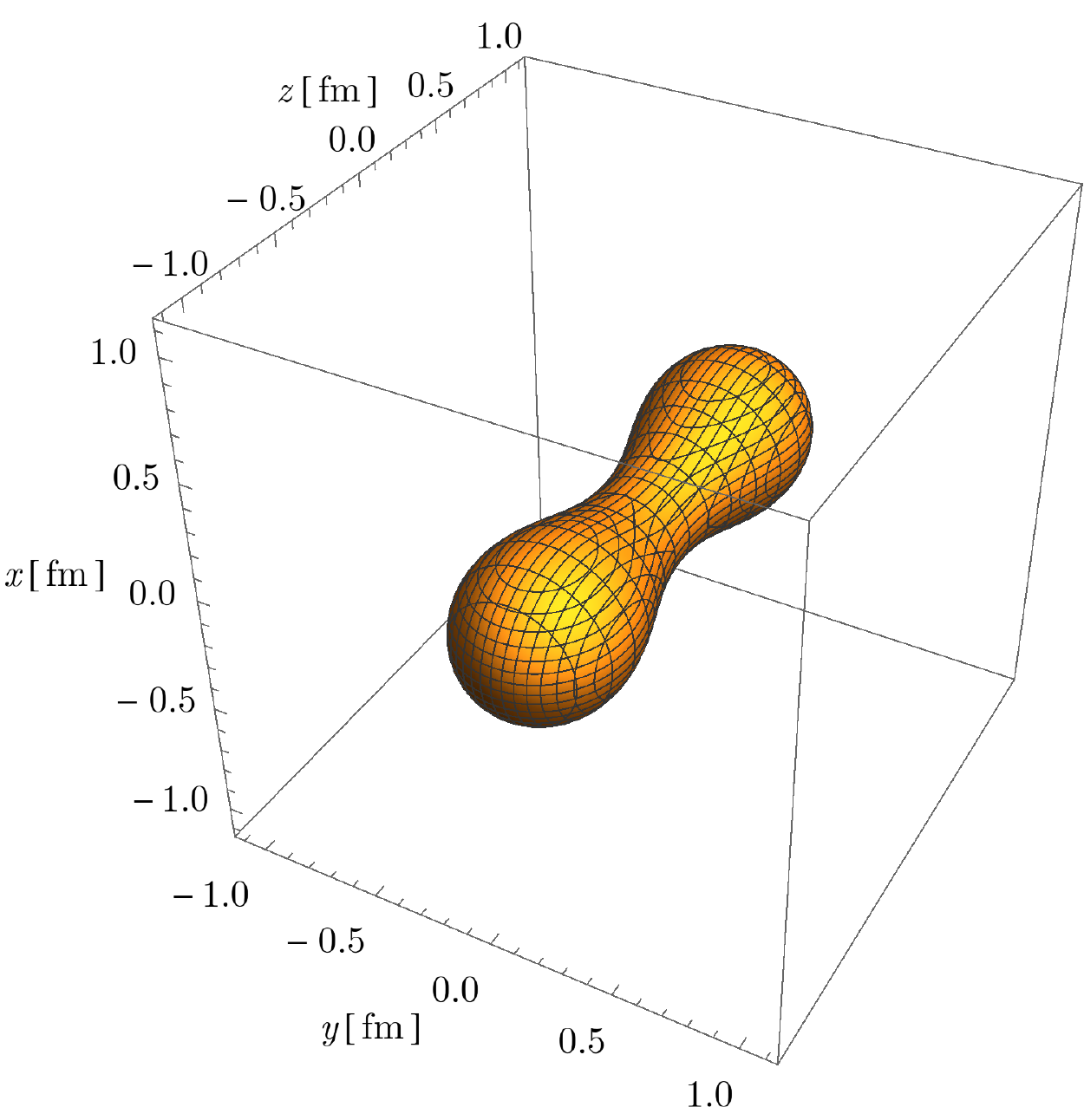}
 \includegraphics[width=0.4\textwidth]{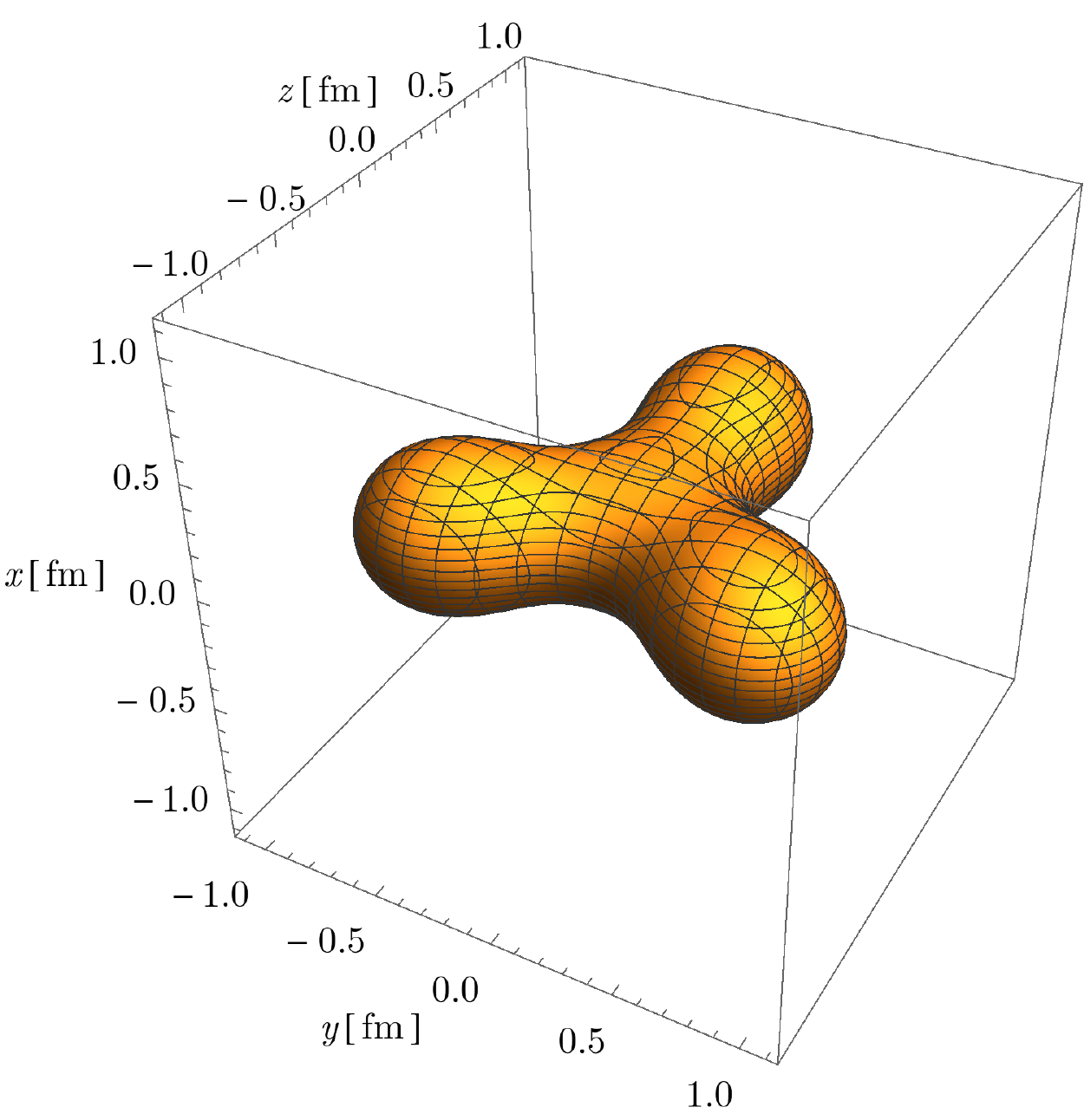}
 \caption{ Effective constituent configurations of typical
 size $s$ in proton: on the left is the quark-diquark 
 configuration labeled in the text as {\bf I} and on 
 the right is the three-quark configuration with a 
 star-junction built from gluons labeled in the text 
 by {\bf Y}.} 
 \label{fig:3}
 \end{figure}

The few-body picture of protons in QCD suggested
by the RGPEP allows us to preliminarily model the 
proton structure in terms of shapes illustrated 
by two typical examples in Fig.~\ref{fig:3}~\cite{KubiczekGlazek}, 
knowing that such models can in future be verified 
in theory. We ask if the ridge effect can phenomenologically
distinguish between the effective configurations. 

We consider three types of configurations. The proton 
quark-diquark configuration, denoted by {\bf I} and 
shown on the left-hand side of Fig.~\ref{fig:3}, is 
motivated by Refs.~\cite{bbg,Bjorken}. It is a superposition 
of a few Gaussians that represent a quark, a diquark and 
gluons forming a tube in between. The three-quark 
configuration, denoted by {\bf Y} and shown on the 
right-hand side of Fig.~\ref{fig:3}, is motivated by 
Ref.~\cite{glazek}. It is a superposition of Gaussians 
that represent three quarks and additional gluons 
forming the Y-shaped junction. The shape of {\bf Y} 
configuration is kept fixed. In addition, we consider
the Gaussian fluctuating three-quark configuration,
denoted by {\bf G-f}, which is the same as the {\bf Y} 
configuration but with the shape parameters generated
according to Gaussian probability distributions.
Details of all configurations we consider are available
in Refs.~\cite{KubiczekGlazek,Kubiczek}. 

%%%%%%%%%%%%%%%%%%%%%%%%%%%%%%%%%%%%%%%%%%%%%%%%%%%%%%%%%%%%%%
\section{ Asymmetries in the initial stage of a $pp$ collision 
          and the final state correlations }
\label{A}
%%%%%%%%%%%%%%%%%%%%%%%%%%%%%%%%%%%%%%%%%%%%%%%%%%%%%%%%%%%%%%

Following Ref.~\cite{dent}, we adapt a simple Glauber 
model, widely used for modelling high energy nuclear collisions \cite{miller}, to describe the density of binary partonic 
collisions in scattering of two systems $A$ and~$B$,
\beq
n_{\mathrm{coll}}(x,y;b,\mathbf{\Sigma}_A, \mathbf{\Sigma}_B) 
\es \sigma_{gg} \int_{-\infty}^{\infty} dz \, 
\rho\left(x - \frac{b}{2},y,z; \mathbf{\Sigma}_A\right) 
\int_{-\infty}^{\infty} dz' \, 
\rho\left(x + \frac{b}{2},y,z';
\mathbf{\Sigma}_B\right) \ , 
\eeq
which is a function of the coordinates $x$ and $y$ 
in the plane transverse to the colliding beams, the
impact parameter $b$ and the varying parameters 
$\mathbf{\Sigma}$ that identify the proton structure and 
its orientation in space. The coefficient $\sigma_{gg}$ 
denotes a parton-parton scattering cross-section, in 
our estimates on the order of 4 mb, and $\rho$ denotes 
the three-dimensional parton distribution described 
in Sec.~\ref{ps}. 

\begin{figure}[ht]
 \centering
 \includegraphics[width=0.43\textwidth]{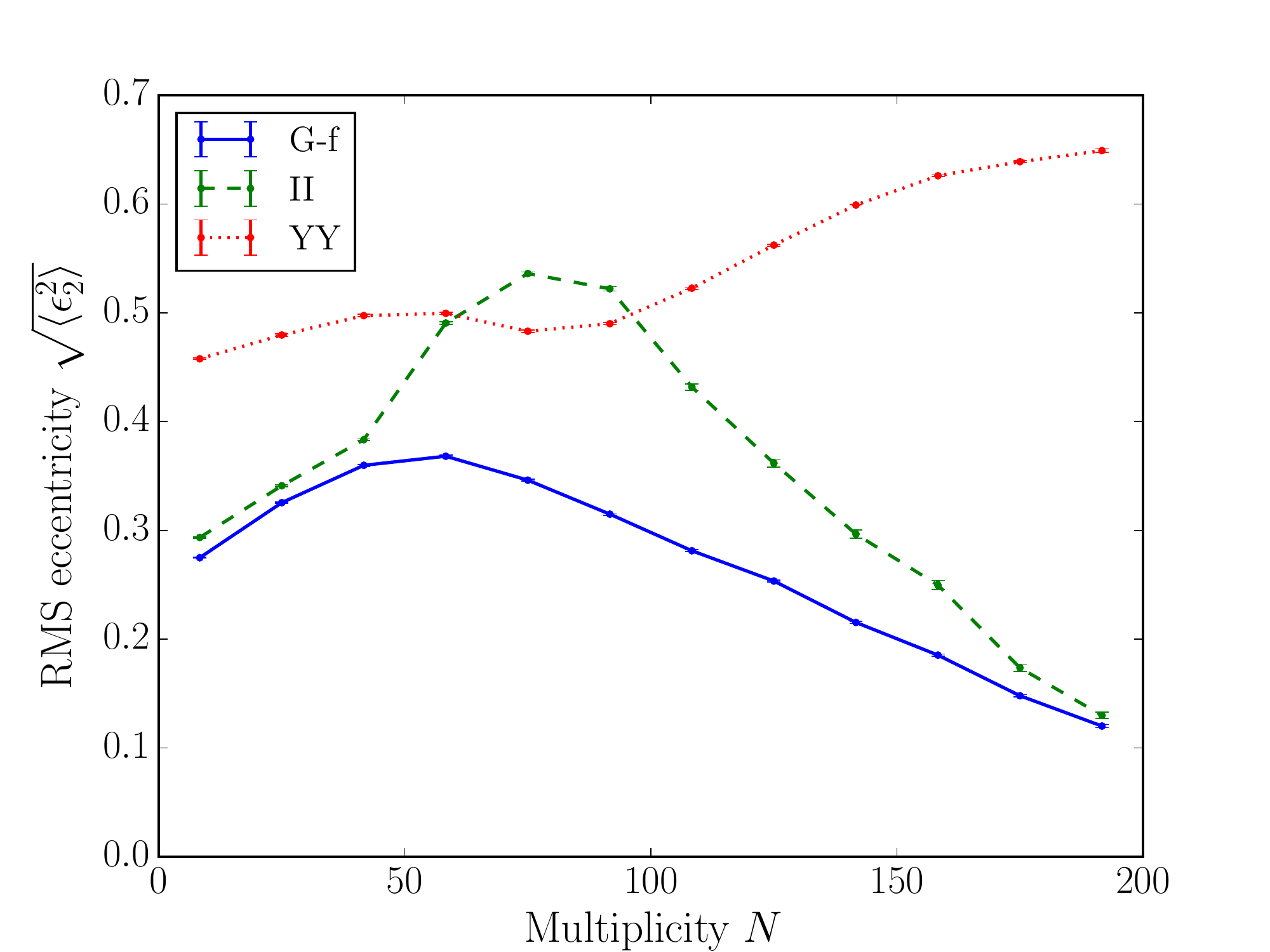}
 \includegraphics[width=0.43\textwidth]{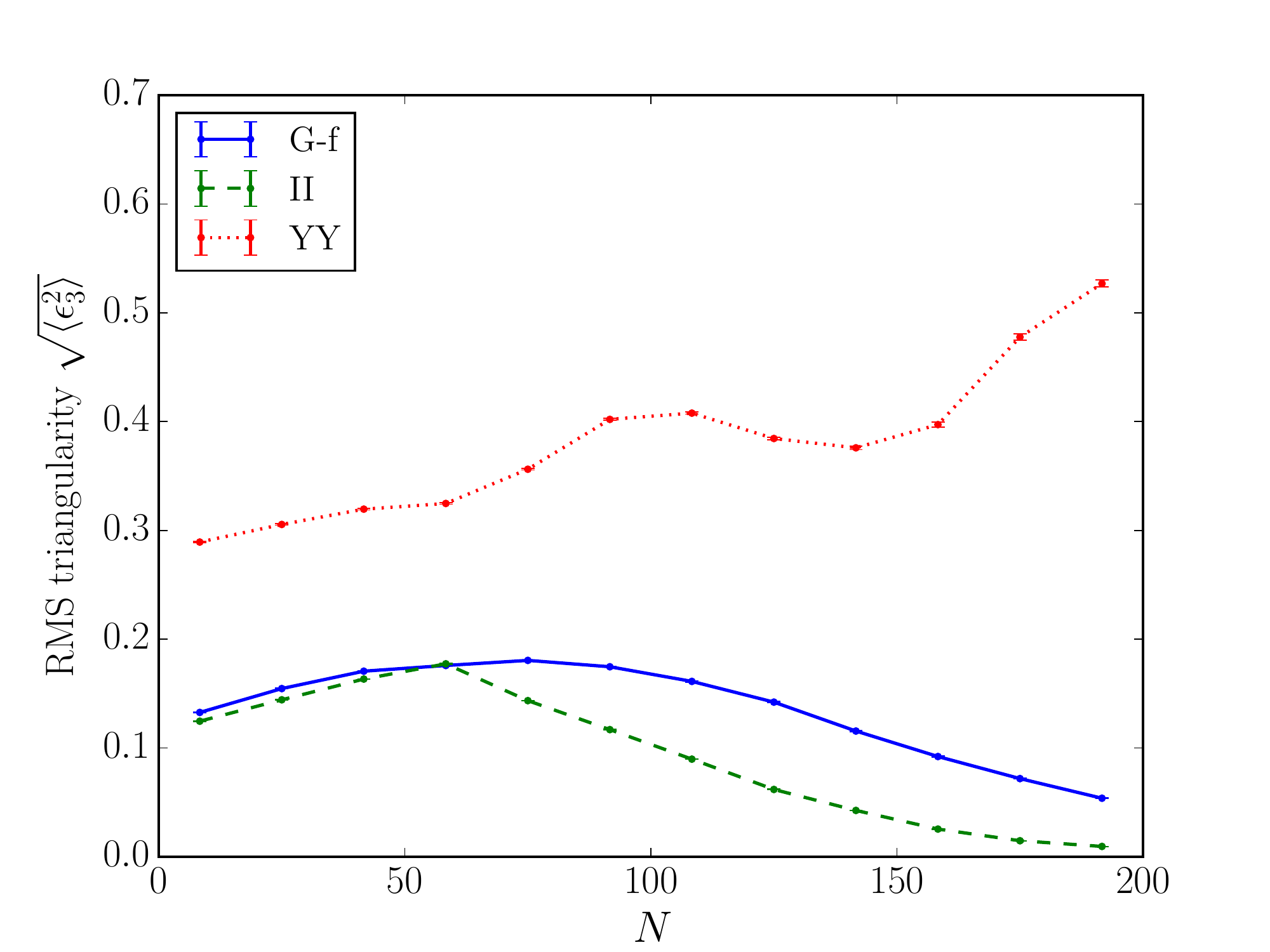}
 \caption{ Average eccentricity and triangularity obtained using 
 Eq.~(\ref{eq:mean}) for three different types of proton structure
 in $pp$ collisions. Green curves labeled {\bf II} correspond
 to collisions of protons in configuration {\bf I} in Fig.~\ref{fig:3},
 and the red curves labeled {\bf YY} correspond to collisions in 
 configurations {\bf Y} in Fig.~\ref{fig:3}. The blue lines labeled 
 {\bf G-f} correspond to the Gaussian fluctuating three-quark 
 configuration in which the {\bf Y}-type proton configurations 
 appear with different shape parameters according to a Gaussian
 probability distribution~\cite{KubiczekGlazek}. Note the distinct 
 multiplicity dependence in the case of {\bf YY} configurations. }   
 \label{fig:4}
 \end{figure}

Eccentricity $\epsilon_2$ and triangularity $\epsilon_3$
in the initial stage of $pp$ collision are calculated
using the formula~\cite{blaizot}
\beq
\label{epsilonn}
\epsilon_n = \frac{\sqrt{\left\{ s^n \cos(n \phi)\right\}^2 
+\left\{ s^n \sin(n \phi)\right\}^2}}{\left\{ s^n
\right\}} \ ,
\eeq
in which the curly brackets denote the expectation value
\beq
\left\{ f(x,y) \right\} \es \frac{\int dx \, dy \ f(x, y) \ 
n_{\mathrm{coll}}(x,y;b,\mathbf{\Sigma}_A,\mathbf{\Sigma}_B)}
{\int dx \, dy \, n_{\mathrm{coll}}(x,y;b,\mathbf{\Sigma}_A, 
\mathbf{\Sigma}_B)} \ ,
\eeq
and coordinates are parameterized as $x = s \cos{\phi}$, 
$y = s \sin{\phi}$. The number of collisions in an event is 
\beq
N_{\mathrm{coll}}(b,\mathbf{\Sigma}_A, \mathbf{\Sigma}_B) 
= \int dx \, dy \, n_{\mathrm{coll}}(x,y;b,\mathbf{\Sigma}_A, 
\mathbf{\Sigma}_B)
\eeq
and the cross-section density in the impact parameter plane is
\beq
\sigma(b,\mathbf{\Sigma}_A, \mathbf{\Sigma}_B) =
1 - \left[ 1 - \frac{N_{\mathrm{coll}}(b,\mathbf{\Sigma}_A, 
\mathbf{\Sigma}_B)}{N_g^2}\right]^{N_g^2} \ . 
\label{eq:cross-section} 
\eeq
The total $pp$ cross-section is thus 
\beq 
\sigma_{pp} \es  \int_0^{\infty} 2\pi b \, db \int 
P(\mathbf{\Sigma}_A) \, d \mathbf{\Sigma}_A \int 
P(\mathbf{\Sigma}_B) \, d \mathbf{\Sigma}_B \
\sigma(b,\mathbf{\Sigma}_A, \mathbf{\Sigma}_B) \ ,
\eeq
where $P(\mathbf{\Sigma})$ is the probability density 
for proton configuration $\mathbf{\Sigma}$. For any
quantity $Q$, its expectation value in many collisions
is
\beq
\left< Q \right> \es \frac{1}{\sigma_{pp}} \int_0^{\infty} 2\pi 
b \, db \int P(\mathbf{\Sigma}_A) \, d \mathbf{\Sigma}_A 
\int P(\mathbf{\Sigma}_B) \, d \mathbf{\Sigma}_B 
\
\sigma(b,\mathbf{\Sigma}_A, \mathbf{\Sigma}_B) \, 
Q(b,\mathbf{\Sigma}_A, \mathbf{\Sigma}_B) \ .
\label{eq:mean} 
\eeq
We used randomly oriented proton configurations in 
the Monte Carlo generation of about $3 \cdot 10^5$ 
events for each proton model and estimated the 
averaged eccentricity $\epsilon_2$ and triangularity 
$\epsilon_3$ in the resulting samples. In our estimates, 
$\sigma_{gg} \sim 4.3$ mb~\cite{v2} and $\sigma_{pp} 
\sim 60$ mb~\cite{cms1} required the number of scattering 
partons $N_g = 9 \pm 2$ to obtain agreement with 
data, assuming that the multiplicity $ N = \alpha 
N_\mathrm{coll}$ and reproducing the value $\langle 
N \rangle = 30$~\cite{cms2} for charged particles by 
choosing $\alpha = 8 \pm 3 $. Our minimum bias  
results for eccentricity and triangularity are shown 
in Fig.~\ref{fig:4} 

It is visible in Fig.~\ref{fig:4} that the initial stage
of $pp$ collision is characterized by different multiplicity
dependence of the asymmetries for different proton structures. 
In collisions of quark-diquark ({\bf II}) and Gaussian-fluctuating 
({\bf G-f}) structures, the asymmetries decrease with multiplicity 
above about $N=100$, while in the collisions of tripod three-quark 
configurations ({\bf YY}) the initial stage asymmetries persist or 
even increase above $N=100$. 

\begin{figure}[ht]
 \centering
 \includegraphics[width=0.8\textwidth]{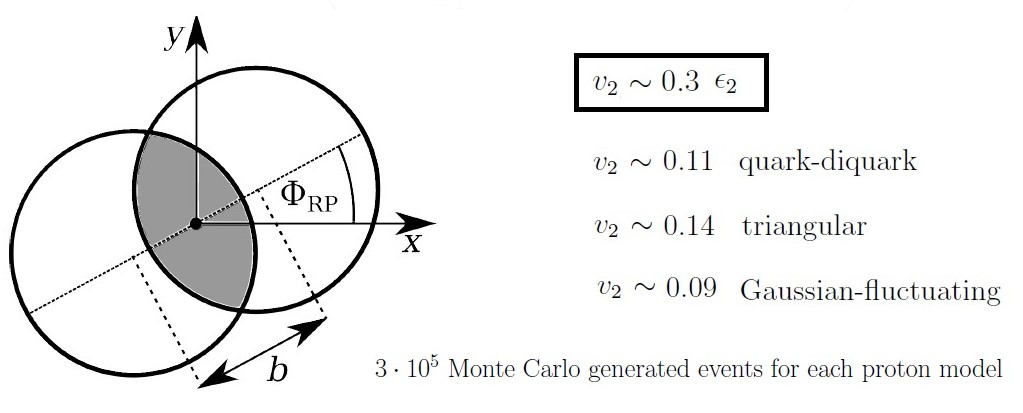}
 \caption{ View of the initial stage of $pp$ collision along the beam 
  for illustration of Eq.~(\ref{Nd}). The average value of the elliptic 
  flow coefficient $v_2(p_T,\eta)$, denoted by $v_2$, is assumed to be 
  proportional to the eccentricity $\sqrt{ \left< \epsilon_2^2 \right>}$ 
  obtained in 
  Eq.~(\ref{eq:mean}) with a coefficient on the order of 0.3~\cite{v2}.
  The magnitudes of $v_2$ resulting from different proton models are
  displayed with quark-diquark = {\bf II}, triangular = {\bf YY}. }
 \label{fig:5} 
 \end{figure}

In order to relate the eccentricity and triangularity 
to data, we note that the observable multiplicity 
distributions in transverse momentum $p_T$ and 
pseudorapidity $\eta$ are conventionally written as
\beq
\label{Nd}
{d^3 N \over d^2p_T d\eta}
\es
\left\{ 1 + 2 \sum_{n=1}^\infty v_n(p_T,\eta) \,
              \cos\left[ n (\phi - \Phi_{RP}) \right] \right\}
\
{ d^2 N \over 2 \pi p_T dp_T d\eta} \ ,
\eeq              
where the reaction plane angle $\Phi_{RP}$ for colliding 
spherically symmetric distributions is illustrated in 
Fig.~\ref{fig:5}. In the actual events the angles 
$\Phi_{RP}$ are determined from the particle 
distributions.

Assuming that the averaged minimal bias elliptic flow parameter $v_2$ is
proportional to the minimal bias eccentricity parameter $\epsilon_2$ with
coefficient order 0.3~\cite{v2}, we obtain $v_2 \sim 0.11$, 0.14 and 0.09 for
the proton models {\bf II}, {\bf YY} and {\bf G-f}, respectively, while 
data indicates $v_2$ in the range 0.04-0.08~\cite{bozek}. 

More generally, taking into account that the initial-stage asymmetry 
parameters $\epsilon_n$ in Eq.~(\ref{epsilonn}) and the final-state 
coefficients $v_n$ in Eq.~(\ref{Nd}) are small, and assuming that 
a set of averaged coefficients $v_n$ with different $n$ depends 
approximately linearly on the set of averaged parameters $\epsilon_n$,
one can infer the dependence of $v_n$ on multiplicity $N$ from the 
dependence of $\epsilon_n$ on $N$. Accordingly, Fig.~\ref{fig:4} 
suggests that the averaged elliptic flow $v_2$ and higher correlations 
in high-energy high-multiplicity $pp$ events may indicate which 
configurations of effective quarks and gluons in the proton structure 
are most likely to occur. Namely, only the {\bf YY} configurations lead 
to $\epsilon_2$ and $\epsilon_3$ that do not fall off for $N$ exceeding 
about 100. Actually, recent data~\cite{ATLAS} from ATLAS Collaboration for $pp$ 
collisions with $\sqrt{s} \sim 13$ TeV show that $v_2$ is stable 
for large $N$. According to our analysis, this finding favors the 
configuration {\bf YY}.

It should be observed that the long-range, near-side angular 
correlations in $pp$ collisions at LHC energies can also be 
studied in terms of multiparton interactions~\cite{inspirehepnetrecord1093441}.
Interactions of four partons to four partons and beyond allow 
for linking the ridge effect to the models~\cite{Broniowski-LC2015}
and theory~\cite{Gaunt-LC2015} of double parton distributions and
their light-front analysis~\cite{Rinaldi-LC2015}. A unified approach
is hence greatly desired.

%%%%%%%%%%%%%%%%%%%%%%
\section{ Conclusion }
%%%%%%%%%%%%%%%%%%%%%%

Simple model estimates suggest that the correlations among final-state 
particles in high-energy high-multiplicity $pp$ collisions are sensitive 
to the proton structure. It is not excluded that future sophisticated 
calculations will identify some spatial features of protons through 
precise interpretation of experimental data on these correlations.
The multidimensional linear relationship between initial-state asymmetries 
and final-state flow parameters that we assume in our estimates requires
verification, e.g., in the hydrodynamic model. If confirmed, it would 
provide an efficient way for studying the proton structure through 
correlations in high-multiplicity $pp$ collisions using the universal
matrices determined by the nature of assumed underlying collision 
dynamics. Interpreting the most recent LHC data on multiplicity 
dependence of the elliptic flow coefficient $v_2$ using our simple 
estimates, suggests that protons may often occur in the configuration 
of three effective quarks connected by a $Y$-shaped gluon junction.

\end{document}